Transport and NQR Studies of Nd$_{1.6-x}$Ce$_x$Sr$_{0.4}$CuO$_4$ with T* Structure

Makoto A$_{MBAI}$[1,2], Yoshiaki K$_{OBAYASHI}$[1,2], Satoshi I$_{IKUBO}$[1] and Masatoshi S$_{ATO}$[1,2]

[1]*Department of Physics, Division of Material Science, Nagoya University, Furo-cho, Chikusa-ku, Nagoya 464-8602*
[2]*CREST, Japan Science and Technology Corporation (JST)*

Abstract

Transport and Cu-NQR studies have been carried out for the T*-type high temperature superconducting Cu oxide Nd$_{1.6-x}$Ce$_x$Sr$_{0.4}$CuO$_4$ with single layered CuO$_2$ planes formed by CuO$_5$ pyramids. No anomaly related to the static or quasi static "stripe" order has been observed in the temperature $T$ and $x$ dependence of the thermoelectric powers $S$. We have not found such kind of anomaly in the $x$ dependence of the superconducting transition temperature $T_c$, either. Although the decrease of the Cu NQR intensity with decreasing $T$ (wipeout) has been observed in this system, it can be understood by considering the loss of the itinerant nature of the electrons, which takes place in the proximity region of the metal-insulator phase boundary, and cannot be connected with the static or quasi static "stripe" order.

Corresponding author: M. Sato (e-mail: e43247a@nucc.cc.nagoya-u.ac.jp)

§1. Introduction

In the study on the mechanism of superconductivity of high-$T_c$ Cu oxides, various kinds of anomalous physical properties have been found. Among these, the pseudo gap can be considered to be closely related to the occurrence of the superconductivity and seems to present key information in describing the detailed features of the electronic behavior.[1-3] The "stripe" ordering found in $La_{2-x-y}R_ySr_xCuO_4$ (R = rare earth elements, $y \neq 0$) (LRSCO), which is now considered to be primarily driven by the charge ordering of the holes and accompanied by the antiferromagnetic ordering of the Cu spins, has introduced another kind of complication to treat the electron system. Actually, the 1/8-anomaly found in various physical quantities including the superconducting transition temperature $T_c$ of $La_{2-x}Ba_xCuO_4$ (LBCO)[4-7] and LRSCO,[8-11] originates from this ordering.[12] The $T_c$- suppression induced by the fluctuation of the "stripes" has been also found in $La_{2-x}Sr_xCuO_4$ in the region around $x =$ 1/8.[13,14] If the "stripes" commonly appear in all high-$T_c$ Cu-oxides, their effects should be considered in the discussion of the mechanism of the superconductivity. It is, therefore, interesting to make clear whether the "stripe" order or its slow fluctuation is confined to $La_{2-x-y}R_yM_xCuO_4$ (M = Sr or Ba) (La214) system or not. It is well-known that effects of the "stripe" order appear in the transport quantities, such as the electrical resistivity $\rho$, Hall coefficient $R_H$ and thermoelectric power $S$.[5,11] In particular, $S$ exhibits a very rapid decrease with decreasing temperature $T$ at the temperature of the transition related to the 1/8 anomaly or related to the "stripe" ordering.[5,11] Moreover the "stripe" order and its slow fluctuation have been pointed out to be able to be detected as the decrease of the Cu-NQR intensity, which is called wipeout,[15,16] even though the decrease is also induced by the loss of itinerant nature of the electrons.[15,17,18]

Several works to search the experimental evidence for the existence of "stripes" have been carried out for $YBa_2Cu_3O_{6+y}$ and $Bi_2Sr_2CaCu_2O_{8+}$ by Akoshima et al.[19] They have reported the anomalous $x$ dependence of $T_c$ and the transport quantities of $Bi_2Sr_2Ca_{1-x}Y_x(Cu_{1-y}Zn_y)_2O_{8+\delta}$, and have pointed out that the anomalies are related to the existence of the "stripes". It has been also reported that anomalous magnetic properties exist at low temperatures in $YBa_2Cu_{3-2y}Zn_{2y}O_{7-\delta}$ with $y = 0.025$ and $7-\delta = 6.65$.[20] On the other hand, no experimental evidence for the "stripe" formation has been found in the superconducting transition temperature $T_c$ and various transport quantities such as $\rho$, $S$ and $R_H$ for $Y_{1-x}Pr_xBa_2Cu_3O_{6+y}$ ($x = 0 \sim 0.5$)[21] and $Bi_{1.7}Pb_{0.3}Sr_2Ca_{1-x}Y_xCu_{2-y}Zn_yO_{8+\delta}$ ($y = 0$ and 0.2).[22] Cu-NQR studies on $YBa_2(Cu_{1-x}Zn_x)_3O_y$ ($x = 0$ and 0.2, $y = 6.55 \sim 6.89$) have not found any evidence for the static or quasi-static "stripe" order, either.[18,23]

In the present paper, further information on the issue if the static "stripe" order or its slow fluctuation commonly exists or not in high-$T_c$ systems is given by presenting results of various kinds of measurements on $Nd_{1.6-x}Ce_xSr_{0.4}CuO_4$,[24,25] whose structural characteristics are different from those of La214 and $YBa_2Cu_3O_{6+y}$ systems: The La214 system consists of the stacking of the single layered $CuO_2$ planes formed by the linkage of $CuO_6$ octahedra, and exhibits structural transition to the low temperature tetragonal (LTT) phase,[9] in which the "stripe" order takes place. $YBa_2Cu_3O_{6+y}$ and $Bi_2Sr_2CaCu_2O_{8+}$ systems have double layered $CuO_2$ planes formed by the linkage of $CuO_5$ pyramids, and exhibit no structural transition. The present compound has single layered $CuO_2$ planes formed by $CuO_5$ pyramids. We report results of transport and Cu-NQR studies on $Nd_{1.6-x}Ce_xSr_{0.4}CuO_4$ and discuss them in comparison with those of La214 system.

§2. Experimental

Samples were prepared by conventional solid state reaction. Mixtures of $Nd_2O_3$, $CeO_2$, $SrCO_3$ and CuO with proper molar ratios were ground and prereacted at 900 ˚C for 12 hours in flowing oxygen gas. The obtained materials were ground again and pressed into pellets, which were fired in flowing oxygen at 1000 ˚C for 24 hours and 1135 ˚C for 24 hours. The x-ray diffraction measurements confirmed that single phase samples of $Nd_{1.6-x}Ce_xSr_{0.4}CuO_{4-\delta}$ ($0.2 \leq x \leq 0.3$) were obtained. In the samples with $x =$ 0.19, slight amounts of impurity phases were found. All these samples were annealed at 500 °C for 60 hours under 26 atm $O_2$ pressure to remove the oxygen deficiency. After this heat treatment, the samples exhibit superconducting transition(see §3). The electrical resistivities $\rho$ were measured by the four probe method with an ac-resistance bridge. Thermoelectric powers $S$ were measured by a dc method, where the typical temperature difference between two ends of samples was 0.2 ~ 2.0 K depending on the temperature region. The correction for the contribution from the Au electrodes was made, where the $S$ values of Au electrodes below ~90 K were obtained by a similar measurement for a $Bi_2Sr_2CaCu_2O_{8+\delta}$ sample in the superconducting state, and for $T>T_c$, values reported by Pearson[26] were used. The Cu NQR spectra and the Cu nuclear spin–spin relaxation time $T_2$ were measured by using a $\pi/2 - \tau - \pi$ spin echo sequence. The $T_2$ measurements were performed at the peak position of the NQR spectra.

§3. Results and Discussion

Figures 1(a) and 1(b) show, respectively, the $x$ dependence of $T_c$ and the $T$-dependence of the shielding diamagnetic moment $M$ measured with the external magnetic field $H$ of 10 Oe for several $x$-values. In the former figure, closed circles are determined by the extrapolation of the steepest part of the $M/H$-$T$ curves in Fig. 1(b) to the background susceptibility curves, which are mainly determined by the Curie-like contribution from the Nd moments. The onset values of $T_c$ shown by the closed triangles are determined from the $M/H$-$T$ curves as the temperatures where $M/H$ deviates from the background susceptibility curves, while the onset values of $T_c$ determined by the $\rho$-$T$ curves and the $S$-$T$ curves in Fig. 2 are shown by dots and open triangles, respectively(Although zero-resistivity has not been observed for the sintered samples because of the possible grain boundary effects, the resistivities exhibit a sharp decrease with decreasing $T$ at temperatures, $T_c$. The existence of the grain boundary effects for the present sintered samples is suggested by the fact that the resistivity of single crystals with no (or less) grain boundaries which were annealed under the same condition as that for the sintered samples exhibits metallic $T$-dependence and zero resisitivity below $T_c$). The offset values of $T_c$ determined as the temperature where $S$ becomes almost zero with decreasing $T$ are also shown by open squares. Although the open triangles determined by the *S-T* data have the largest values because of the reasons that *S* is sensitive to the filamentary superconductivity and insensitive to the grain boundary effects, the hole-concentration (*p*) dependences are very similar for all kinds of the $T_c$ data. Here, we draw the solid line as the guide for the eye by using the closed circles, because the *M/H* data reflects the transition to the bulk superconductiviting phase, and can, we think, present the most reliable *x*-dependence.

The maximum value of the observed $T_c$ is about 22 K at $x \sim 0.20$. The actual *p* value may be smaller than the nominal value ($0.4 - x$) at least by ~0.05, because of the existence of the oxygen deficiency. The so-called 1/8-anomaly of $T_c$ cannot be found in the $T_c$ -*x* curve.

The thermoelectric power *S* shown in Fig. 2 exhibits the similar *T*-dependence to that observed for other high-$T_c$ systems.[27] The value of *S* at 300 K smoothly increases with increasing *x*, indicating the non-existence of 1/8 anomaly in its *x*-dependence. We have not found any anomaly in the *T*-dependence of *S*, either. The result should be contrasted to the case of LBCO and LRSCO,[5,11] where the very sharp decrease of *S* has been observed with decreasing *T* in relation to the 1/8 anomaly. Thus, the static "stripe" order does not exist in the present system. The present data of *S* are consistent with those measured for *x*=0.2 by Ikegawa *et al*.[28]

Next we describe the Cu-NQR studies. Examples of Cu NQR spectra of $Nd_{1.6-x}Ce_xSr_{0.4}CuO_4$ with $x = 0.20, 0.24, 0.26, 0.28$ are shown in Fig. 3. The spectra can essentially be fitted by using the superposition of two pairs of $^{63}Cu$ and $^{65}Cu$ Gaussian-lines. But in the figure we show the results, by the thick line, obtained by fitting with three pairs, one of which has only a small intensity. The components which correspond to the $^{63}Cu$ and $^{65}Cu$ nuclei are shown by the thin and broken lines, respectively. We have not carried out the site assignment for these pairs, because in the present study only the total intensity of these pairs obtained by integrating all the observed spectra is discussed. In the temperature region studied here (below 200 K), the line width and the peak position (corresponding to the nuclear quadrupole resonance frequency $\nu_Q$) of each component do not strikingly change with *T*, which is contrasted with the large change of $\nu_Q$ found below the "stripe" ordering temperature in $La_{1.6-x}Nd_{0.4}Sr_xCuO_4$.

The total intensity *I* of the Cu-NQR spectra multiplied by *T*, which is proportional to the number of the observed nuclei, is shown in Fig. 4 as a function of *T* for the samples with $x = 0.20, 0.24, 0.26$ and 0.28. Corrections were made by estimating the intensities at $\tau = 0$ using the transverse relaxation curves of the nuclear spin echo. For all samples, $I \times T$ begins to decrease with decreasing *T* at around $T_{onset}$ which is defined for each sample as the temperature where the qualitative change of the spin-echo decay curve occurs from the mixture of the Gaussian and Lorentzian type decays to the Lorentzian type one (The value of $T_{onset}$ increases monotonically with decreasing *x* or with decreasing *p*). The result indicates that the wipeout is observed for all samples, though the wipeout fraction *F* is smaller than unity.

By comparing the data with those of other systems, we realize that the behavior of the present data of the wipeout is different from that of La214systems: For all the samples studied here, the increase of the fraction *F* with decreasing *T* is very gradual similarly to the case of $YBa_2(Cu_{0.8}Zn_{0.2})_3O_y$ for which the observed wipeout is not, we think, related to the "stripe" order but related to the loss of the itinerant nature of the electrons.[18] In contrast to this, the wipeout fraction $F$ of La214 systems exhibits quite characteristic behavior: $F$ of LRSCO (R=Nd) reported in the region of $p$ larger than 1/8, exhibits an abrupt increase just below the $T_{onset}$ due to the occurrence of the static or quasi "stripe" order,[15, 16] and as was stated in refs. 15 and 16, the $T$-dependence of $F$ is similar to that of the order parameter of the "stripes". The behavior is also found in $La_{2-x}Sr_xCuO_4$ in the narrow region of $p$ close to 1/8(In the region of larger values of $p$, the wipeout does not exist).These results indicate that the wipeout appears with decreasing $T$, as the result of the occurrence of the static or quasi static "stripe" order. Here, the "quasi static" order means that the fluctuation of the "stripe" pattern is very slow and its characteristic time is longer than the inverse NQR frequency. Similarly, it also indicates, in other kinds of experimental studies, that the characteristic time of the fluctuation is longer than the time scale relevant to observed physical quantities. On the other hand, in the region of $p$ smaller than and apart from 1/8, $T_{onset}$ of LRSCO(R=Nd) increases with decreasing $p$ and becomes much larger than the temperature of the "stripe" order, which decreases with $p$. In that $p$ region, the $T$-dependence is gradual even just below $T_{onset}$. These facts indicates that the onset of the wipeout phenomenon is not necessarily related to the (quasi) static "stripe" order. On this point, we will argue once more below how the gradual $T$-dependence observed in $Nd_{1.6-x}Ce_xSr_{0.4}CuO_4$ as well as $YBa_2(Cu_{0.8}Zn_{0.2})_3O_y$[18] originates from the loss of the itinerant nature of the electrons.

As has been already noted in refs. 15,16,18 and 23, the nuclear spin-spin relaxation curves above $T_{onset}$ can be expressed by the relation $m(\tau) = m(0) \cdot \exp(-2\tau/T_{2L} - 1/2 \cdot (2\tau/T_{2G})^2)$, where $m$ is the spin echo amplitude. It changes to the only Lorenzian type one as $T$ decreases through $T_{onset}$. The temperature dependence of $1/T_{2L}$ and $1/T_{2G}$ are shown in Fig. 5. The lowest temperatures at which the Gaussian decay component can be observed, correspond to the values of $T_{onset}$. They are indicated by the arrows in the figure. Above $T_{onset}$, $1/T_{2L}$ of each sample increases only gradually with decreasing $T$, and at ~$T_{onset}$ the rate $d(1/T_{2L})/dT$ exhibits the rather sharp increase, which can be attributed, as Hunt *et al.* pointed out, to the appearance of the additional Lorentzian contribution to the relaxation as well as the disappearance of the Gaussian one.

The magnitude and the $T$ dependence of $1/T_{2G}$ do not sensitively depend on $x$ (or $p$). The increase of the quantities, $T_{onset}$, $F$ and $1/T_{2L}$ (in all $T$ range) with increasing $x$(or with decreasing $p$), cannot be explained by only the magnetic interaction with the Nd moments, because the number of Nd atoms decreases with increasing $x$.

As has been mentioned above, $F$ exhibits only a gradual increase with decreasing $T$ for all the samples studied here. The behavior is in clear contrast to the results observed for the samples of $La_{2-x}Sr_xCuO_4$ and LRSCO (R=Nd,) with $p \geq 1/8$, where $F$ increases abruptly just below the $T_{onset}$. Moreover, the fraction $F$ of the present system increases with increasing $x$ (or with decreasing $p$) in all $T$ and $x$ regions where nonzero $F$ is observed. These results cannot easily be understood by adopting the picture of the "stripe" order, because effects of the "stripe" order are strong at $x$=1/8 as compared with the surrounding $x$ region. Actually, the effects on the electronic physical quantities observed in La$_{1.6-x}$Nd$_{0.4}$Sr$_x$CuO$_4$ with stripe order is the strongest at $x=1/8$,[29]) and in the wipeout fraction $F$ of the system, a local maximum has been observed in a certain region of $T$ as a function of $p$ at $p=1/8$.[16]) It has not been observed in the present measurements.

To understand this result, we consider effects of the electron localization, which becomes significant as $T$ decreases and as $x$ increases (or $p$ decreases). The experimental results on the present system as well as other systems,[15,16,18]) indicate that the wipeout fraction $F$ and $T_{onset}$ increases as the systems approach the metal insulator phase boundary, where the electrons begin to exhibit the tendency of the localization. In systems which have electrons with localized nature, we can expect the inhomogeneous distribution of charges. We can also expect the localized spins induced by this inhomogeneity. The slow fluctuation of these charges and spins can be considered to be responsible for the increase of $T_{onset}$, $F$ and $1/T_{2L}$. Then, it is natural to attribute the wipeout observed here to the loss of the itinerant nature of the electrons.

The electron localization or the loss of the itinerant nature is more significant for systems with larger randomness. We think that the randomness in Nd$_{1.6-x}$Ce$_x$Sr$_{0.4}$CuO$_4$ is large as compared with other high $T_c$ oxides, which may explain the relatively low $T_c$ and the presently observed result that the wipeout fraction does not vanish even for the sample with $p$ as large as ~0.15.

In the above arguments, we have shown that no experimental evidence for the existence of the static or quasi static "stripe" order has been observed in Nd$_{1.6-x}$Ce$_x$Sr$_{0.4}$CuO$_4$ with T*structure. In La$_{2-x}$Ba$_x$CuO$_4$ and La$_{2-x-y}$R$_y$Sr$_x$CuO$_4$ ($y \neq 0$), which exhibit the static "stripe" order, the structural transition to the LTT phase takes place. La$_{2-x}$Sr$_x$CuO$_4$, which has the slight $T_c$-suppression around $x = 1/8$ has the structural fluctuation related to the transition to the LTT phase.[30]) Then, the lattice distortion may have an effect to stabilize the "stripes", and the structural characteristics of the high-$T_c$ oxides should be considered to be important for the existence/nonexistence of the static or quasi static "stripe" order.

§4. Summary

The transport properties and Cu-NQR spectra have been studied for Nd$_{1.6-x}$Ce$_x$Sr$_{0.4}$CuO$_4$ with T* structure. In the $T$ and $x$ dependences of $S$ and in the $x$ dependence of $T_c$, no anomaly related to the static or quasi-static "stripe" order has been observed. The NQR wipeout phenomenon has been observed in this system at low temperatures, though the fraction does not reach unity. Detailed studies on the $x$- and $T$-dependence of the wipeout fraction indicate that the phenomenon can be understood by the loss of the itinerant nature of the electrons and does not indicate the existence of the static or quasi static "stripes".

Figure captions

Fig. 1 (a) $T_c$ values determined by the extrapolation of the steepest part of $M/H - T$ curves shown in (b) to the background susceptibility curves are shown by the closed circles as a function of the Ce concentration $x$. The solid curve is drawn for the data as the guide for the eye. Open and closed triangles and the dots are the onset $T_c$ values determined by measuring the $T$-dependence of $S$, $M/H$ and $\rho$, respectively, while the open squares indicate the offset $T_c$ values determined by the $S$-data. (b) Temperature dependence of the shielding diamagnetic moments $M$ divided by the applied magnetic field $H$ measured in the zero field cooling condition, are shown for various $x$ values.

Fig. 2 Upper panel shows the $x$ dependence of the thermoelectric power $S$ of $Nd_{1.6-x}Ce_xSr_{0.4}CuO_4$ at 300 K. In the lower panel the temperature dependence of $S$ of $Nd_{1.6-x}Ce_xSr_{0.4}CuO_4$ is shown for various $x$ values.

Fig. 3 Cu NQR spectra of the samples of $Nd_{1.6-x}Ce_xSr_{0.4}CuO_4$ with various $x$ values. Thick lines show the results of the fittings to the observed data by using the superposition of Gaussian curves of three sets of $^{63}Cu$ and $^{65}Cu$ spectra.

Fig. 4 Temperature dependence of the Cu NQR intensities multiplied by $T$ are shown for $Nd_{1.6-x}Ce_xSr_{0.4}CuO_4$ samples with various $x$ values. Each line is guide for the eye.

Fig. 5 Temperature dependence of $1/T_{2L}$ and $1/T_{2G}$ estimated by fitting the relation $m(\tau) = m(0)\cdot\exp(-2\tau/T_{2L}-1/2\cdot(1/T_{2G})^2)$ to the observed transverse relaxation curves of the spin echo amplitude $m(\tau)$. The arrows indicate $T_{onset}$, the lowest temperatures where the Gaussian component could be observed. The numbers attached to the arrows indicate the corresponding $x$ values. The solid and broken lines are guides for the eye.

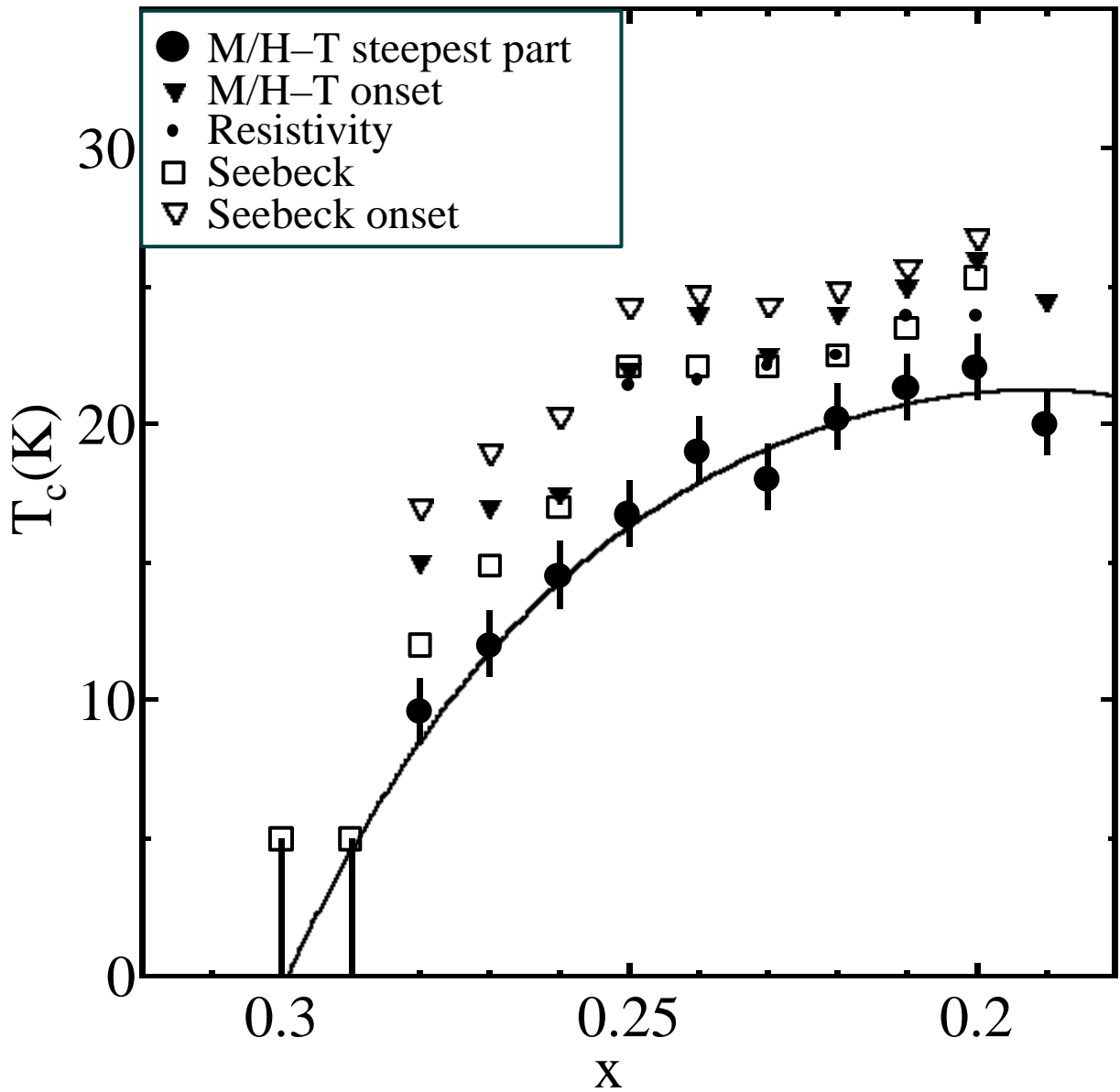

Fig. 1(a)

M. Ambai *et al.*

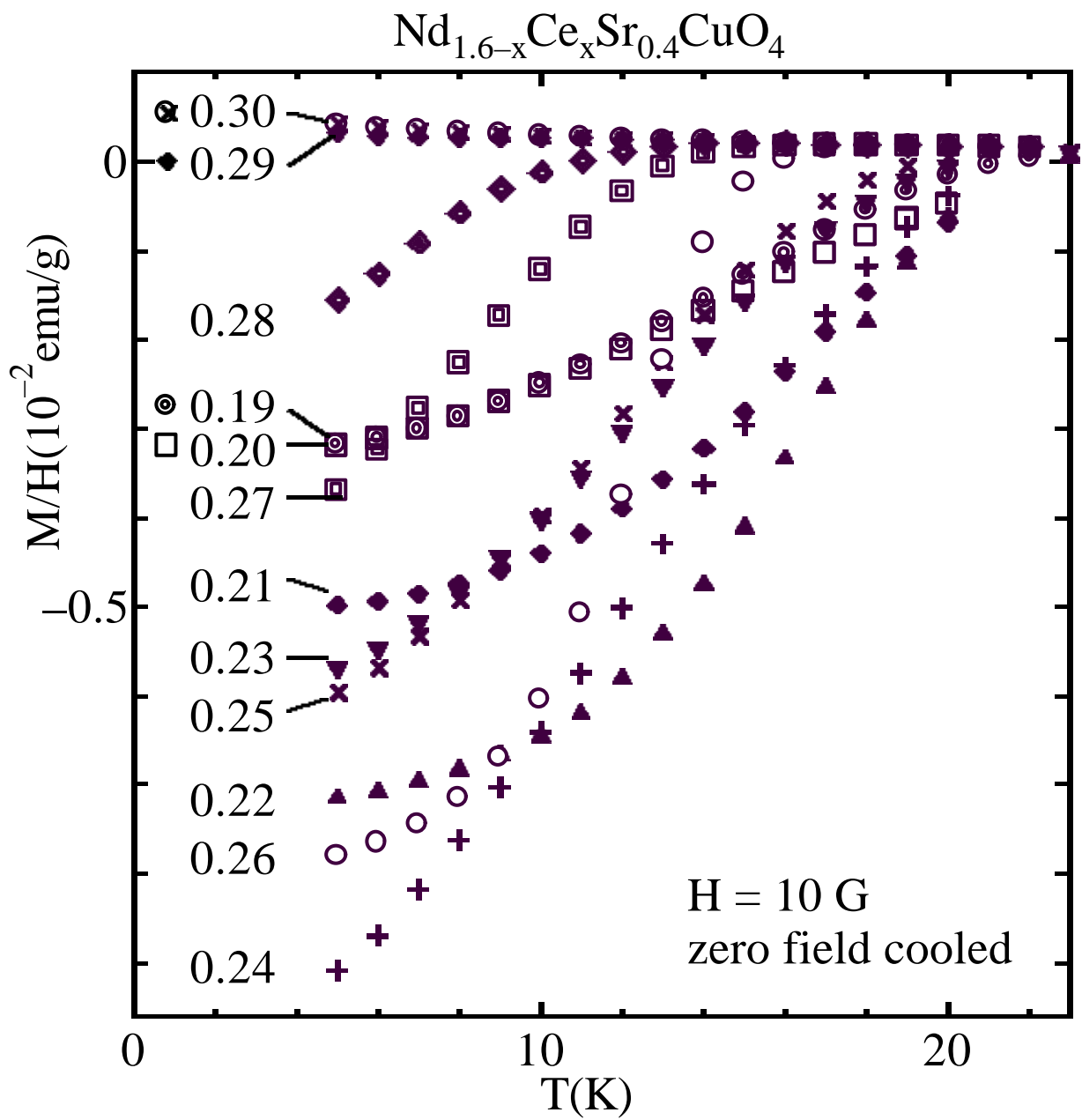

Fig. 1(b)

M. Ambai *et al.*

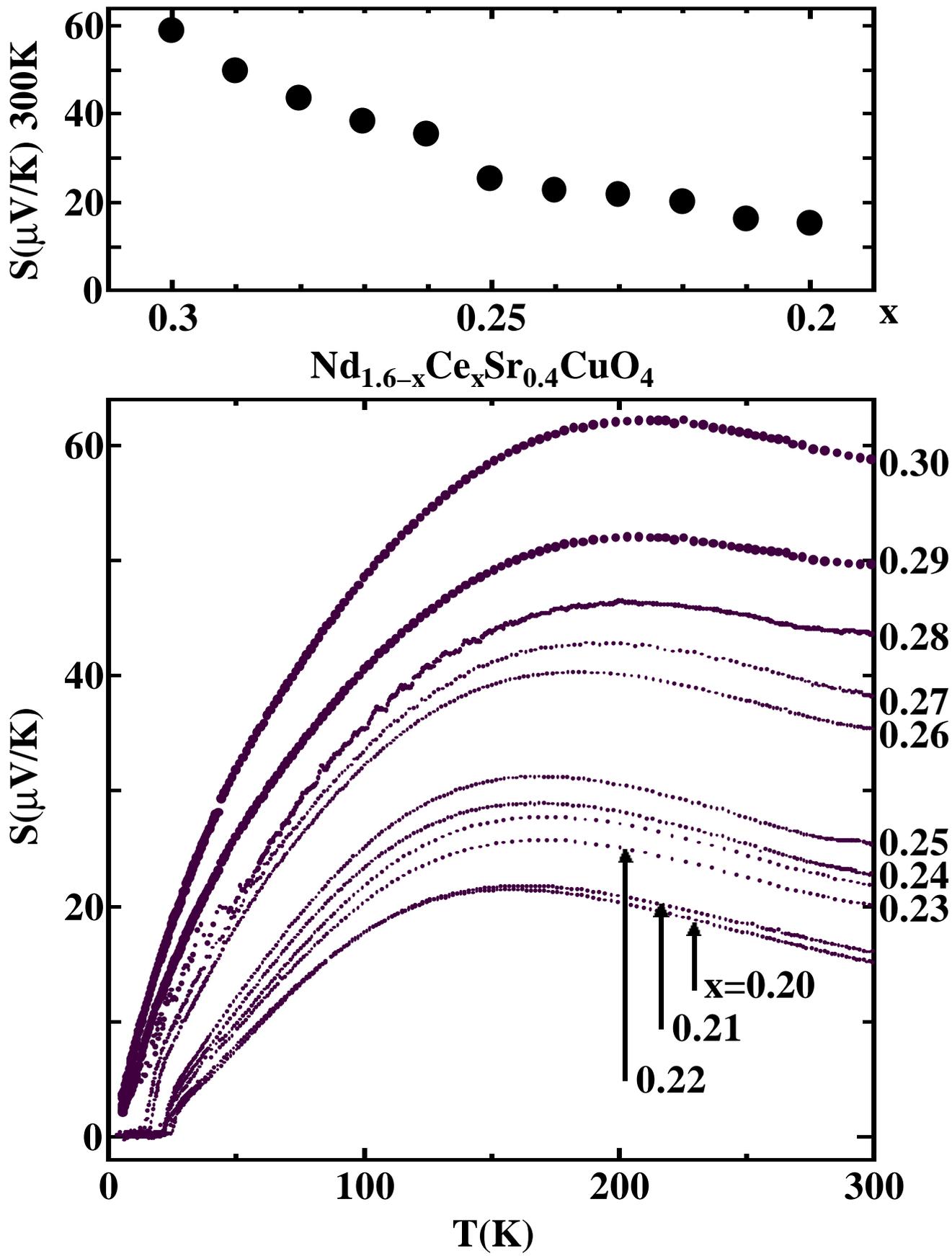

Fig. 2

M. Ambai *et al.*

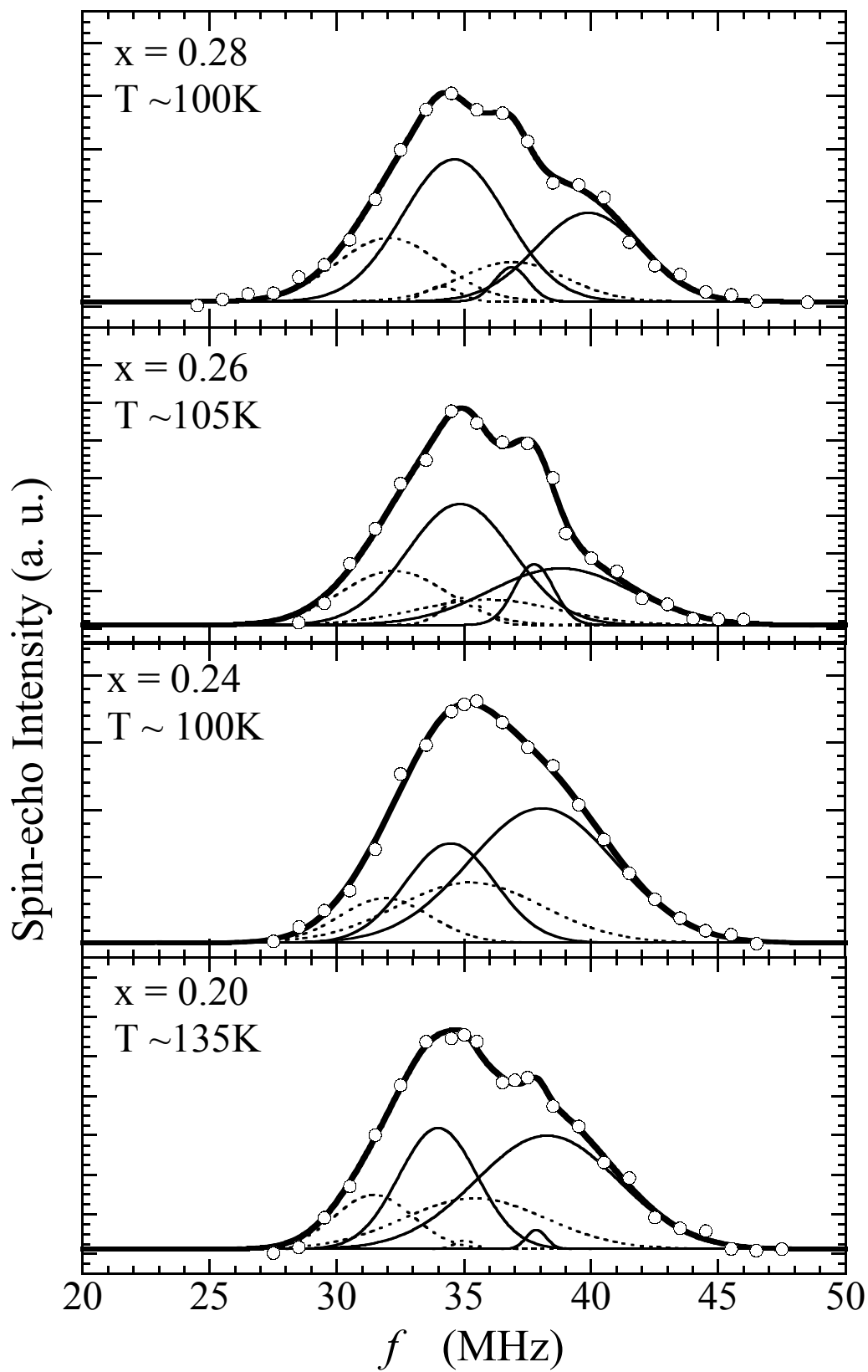

Fig. 3

M. Ambai *et al.*

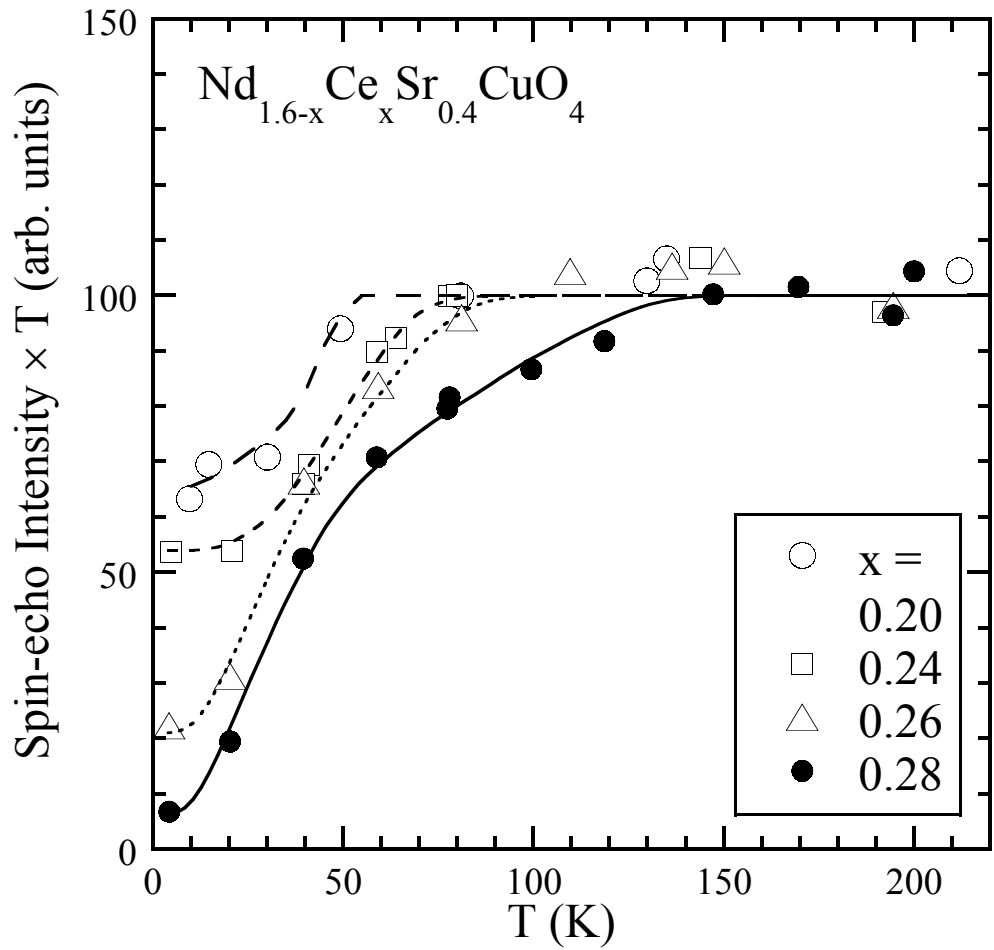

Fig. 4

M. Ambai *et al.*

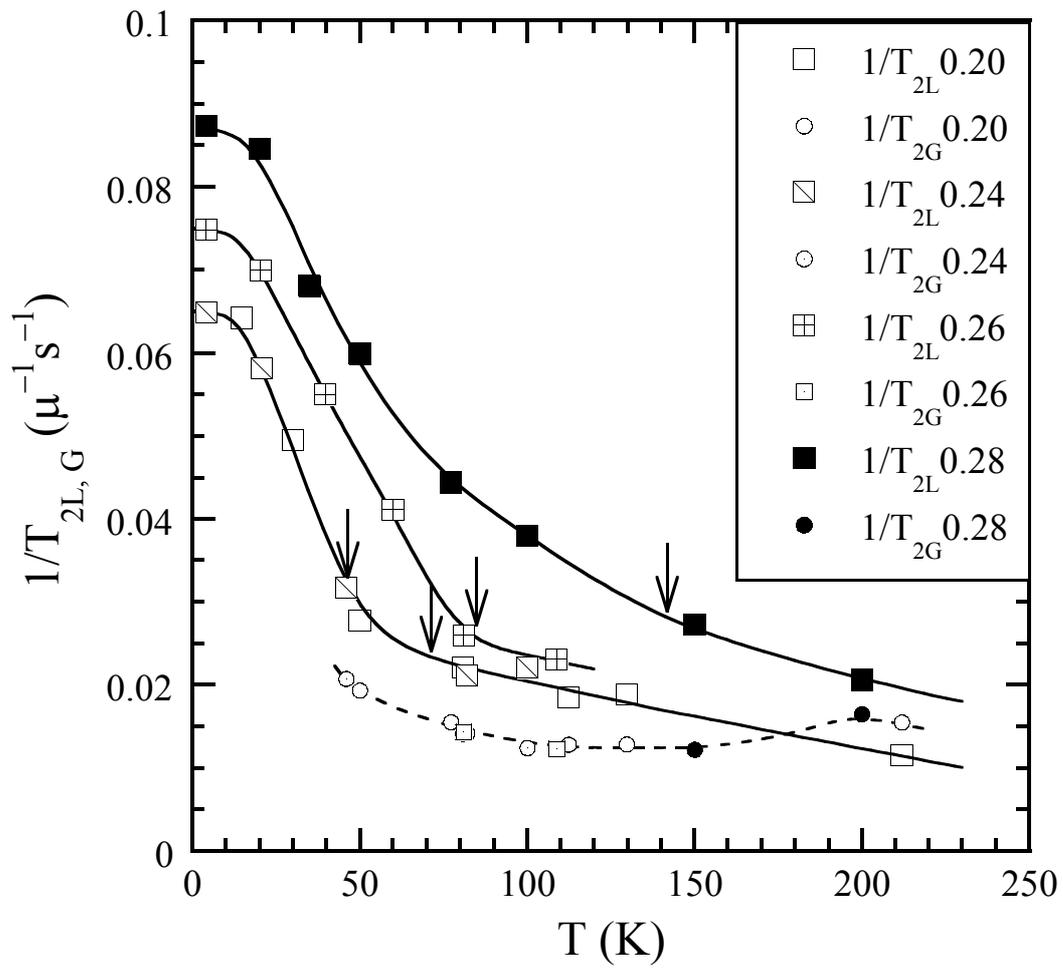

Fig. 5

M. Ambai *et al.*